\documentclass[aps,reprint,groupedaddress,superscriptaddress]{revtex4-1}
\usepackage{amsmath}
\usepackage{physics}
\usepackage{graphicx}
\usepackage{hyperref}
\usepackage{times}
\usepackage{xcolor}
\usepackage{siunitx}
\usepackage{layouts} 

\newcommand{\deltal}{\ensuremath{\delta_{\ell}}} 
\newcommand{\lambdal}{\lambda_{\ell}} 
\newcommand{\wcav}{\omega_c} 
\newcommand{\watom}{\omega_a} 
\newcommand{\deltac}{\delta_c} 
\newcommand{\tramp}{\tau_{\mathrm{ramp}}} 
\newcommand{\ncav}{n(z')} 
\newcommand{\lt}{\ell_t} 
\newcommand{\taul}{\tau_{\ell}} 
\newcommand{\us}{\ensuremath{~\mathrm{\mu s}}}

\newcommand{\nm}{\ensuremath{~\mathrm{nm}}}
\newcommand{\uK}{\ensuremath{~\mu\mathrm{K}}}
\newcommand{\kHz}{\ensuremath{~\mathrm{kHz}}}
\newcommand{\MHz}{\ensuremath{~\mathrm{MHz}}}
\newcommand{\mHz}{\ensuremath{~\mathrm{mHz}}}
\newcommand{\srs}{\ensuremath{^{87}\mathrm{Sr}\,}}
\newcommand{\sre}{\ensuremath{^{88}\mathrm{Sr}\,}}

\newcommand{\subref}[2]{\hyperref[#1]{\ref*{#1}(#2)}}
\definecolor{refblue}{HTML}{2E2E91}
\hypersetup{colorlinks=true, citecolor=refblue, urlcolor=refblue, linkcolor=refblue}

\definecolor{cadmiumgreen}{rgb}{0.0, 0.42, 0.24}

\newcommand{\Rev}[1]{{\color{black}{#1}}}

\begin{document}

\title{Continuous collective strong coupling between atoms and a high finesse optical cavity}

\author{Julia R. K. Cline}
\affiliation{JILA, NIST, and Dept. of Physics, University of Colorado, 440 UCB, Boulder, CO 80309, USA}
\author{Vera M. Schäfer}
\affiliation{JILA, NIST, and Dept. of Physics, University of Colorado, 440 UCB, Boulder, CO 80309, USA}
\author{Zhijing Niu}
\affiliation{JILA, NIST, and Dept. of Physics, University of Colorado, 440 UCB, Boulder, CO 80309, USA}
\author{Dylan J. Young}
\affiliation{JILA, NIST, and Dept. of Physics, University of Colorado, 440 UCB, Boulder, CO 80309, USA}
\author{Tai Hyun Yoon}
\affiliation{Department of Physics, Korea University, Seoul 02841, Republic of Korea}
\author{James K. Thompson}
\affiliation{JILA, NIST, and Dept. of Physics, University of Colorado, 440 UCB, Boulder, CO 80309, USA}

\date{\today}

\begin{abstract}

\Rev{We demonstrate continuous loading of strontium atoms into a high finesse ring cavity and observe continuous strong collective coupling in the form of a vacuum Rabi splitting between the atoms and the cavity on the 7.5~kHz transition $^1$S$_0$ to $^3$P$_1$.  
The atoms are loaded into the cavity from a 3D narrow linewidth molasses, thus avoiding large magnetic field gradients and associated broadening of transition frequencies. 
The ring cavity allows us to realize a deterministic conveyor belt to transport atoms away from the loading region where the laser cooling beams lead to broadening of the strontium clock transition. 
We trap up to $10^6$ atoms in an intracavity 813~nm lattice in the Lamb-Dicke regime, and transport the atoms along the cavity axis. 
This work opens the path to the creation of a continuous wave superradiant laser with millihertz linewidth enabling searches for new physics and the use of high-precision optical frequency references outside of low vibration laboratory environments.}

\end{abstract}


\maketitle 

Laser-cooled and trapped atoms are a powerful tool for quantum metrology \cite{Lange2021,Bothwell2022,McGrew2018,Takamoto2020,Brewer2019,Zheng2022}, quantum information \cite{Srinivas2021a,Ballance2016,Ma2022,Jenkins2022,Graham2022} and quantum sensing \cite{Kasevich1991,Greve2022,Gustavson1997,Peters2001,Norcia_GW2017}, offering excellent precision and sensitivities.
Applications include the study of quantum many-body physics \cite{Martin2013,Zhang19092014} and testing new theories of fundamental physics \cite{Derevianko2014,Derevianko2016,Safronova2017,Castro-Ruiz2020}.
Laser-cooled atoms have been combined with high-finesse optical cavities to realize powerful platforms for exploring many-body physics \cite{Baumann_2010,Georges2018,Schuster2020,muniz2020,holzinger2022}, and creating highly entangled states \cite{Cox2016,Hosten2016,Braverman2019} that will enable next generation quantum sensors \cite{Pedrozo-Penafiel2020,Greve2022}.
The cavity serves as a common communication bus that allows the atoms to interact with each other coherently or dissipatively \cite{Norcia2016b,Norcia_SS_2018}.

The vast majority of laser-cooled neutral atom experiments rely on cyclic loading of atoms, where the atoms are prepared by sequentially applied cooling and trapping stages to reach temperatures sufficiently low to allow for high-precision measurements.
This typically leads to cycle times of 1~s to a few minutes.
There is much interest in providing atoms in a continuous way \cite{Chen2022}, for example in optical tweezers \cite{Singh2022}, continuous clock measurements \cite{Dorscher2020} and matter wave interferometers \cite{Savoie2019}.
A continuous supply of atoms increases the sampling time and duty cycle rate, and therefore improves signal-to-noise ratios and clock stability \cite{Katori2021}, reduces dead-time and aliasing \cite{Kwolek2020}, and gives access to new physical regimes for probing physics.

In this Letter, we demonstrate continuous loading of $2.1(3)\cross 10^7$ $^{88}$Sr atoms/s into a high-finesse ring cavity in the strong collective atom-cavity coupling regime, and subsequent transport of the atoms in a moving intracavity lattice to a region free of laser light where the atoms experience lower decoherence.
Strontium is an excellent choice for continuous loading as it offers cooling transitions of widely differing linewidths.
It has recently been used to realize a continuous Bose-Einstein condensate \cite{Chen2022}.
BECs have been loaded cyclicly into high-finesse ring cavities previously to explore self-organization physics including collective atomic recoil lasing and super-solids \cite{Nagorny2003,Slama2007,Georges2018,Schuster2020}.
To our knowledge, this is the first demonstration of continuous strong atom-cavity coupling in a ring cavity combined with deterministic transport of atoms within the cavity using a moving optical lattice.   

A particular goal of the system presented here is the realization of a continuous superradiant laser \cite{Kleppner1962,Benmessai2008,Chen2009a,MYC09,bohnet2012steady,Maier2014,norcia2015cold,norcia2016superradiance,Kazakov2021,Zhang2021}.
This would be an enormous advance compared to previous work realizing a pulsed superradiant laser on the millihertz optical clock transition in strontium \cite{Norcia2016b,Norcia_SRFreq_2018} which already achieved an observed single-pulse fractional frequency imprecision as low as $6.7(1) \times 10^{-16}$.
The continuous, high-flux source that we demonstrate here should be sufficient to realize a superradiant laser that would eliminate the Dick noise-aliasing effect in passively probed clocks and be $>10^5$ times less sensitive \cite{bohnet2012steady,Norcia_SRFreq_2018} to both technical and thermal vibration noise that limits today's best narrow-linewidth lasers.
This opens a path to higher precision and possibly ultranarrow laser sources that can operate outside of quiet laser laboratories. 
The system would also be ideal for other continuous cavity-enhanced techniques such as magnetically-induced transparency \cite{Winchester2017} or non-linear spectroscopy \cite{Martin_2011,Thomsen2015}.
Furthermore, our setup is an ideal source for a future atom laser, for realizing entangled states for optical lattice clocks \cite{Pedrozo-Penafiel2020} and for inertial sensors due to the uniform atom-cavity coupling \cite{Greve2022,Hu_2017,Salvi2018,shankar2019}.


\begin{figure}[!htb]
\includegraphics[width=3.375in]{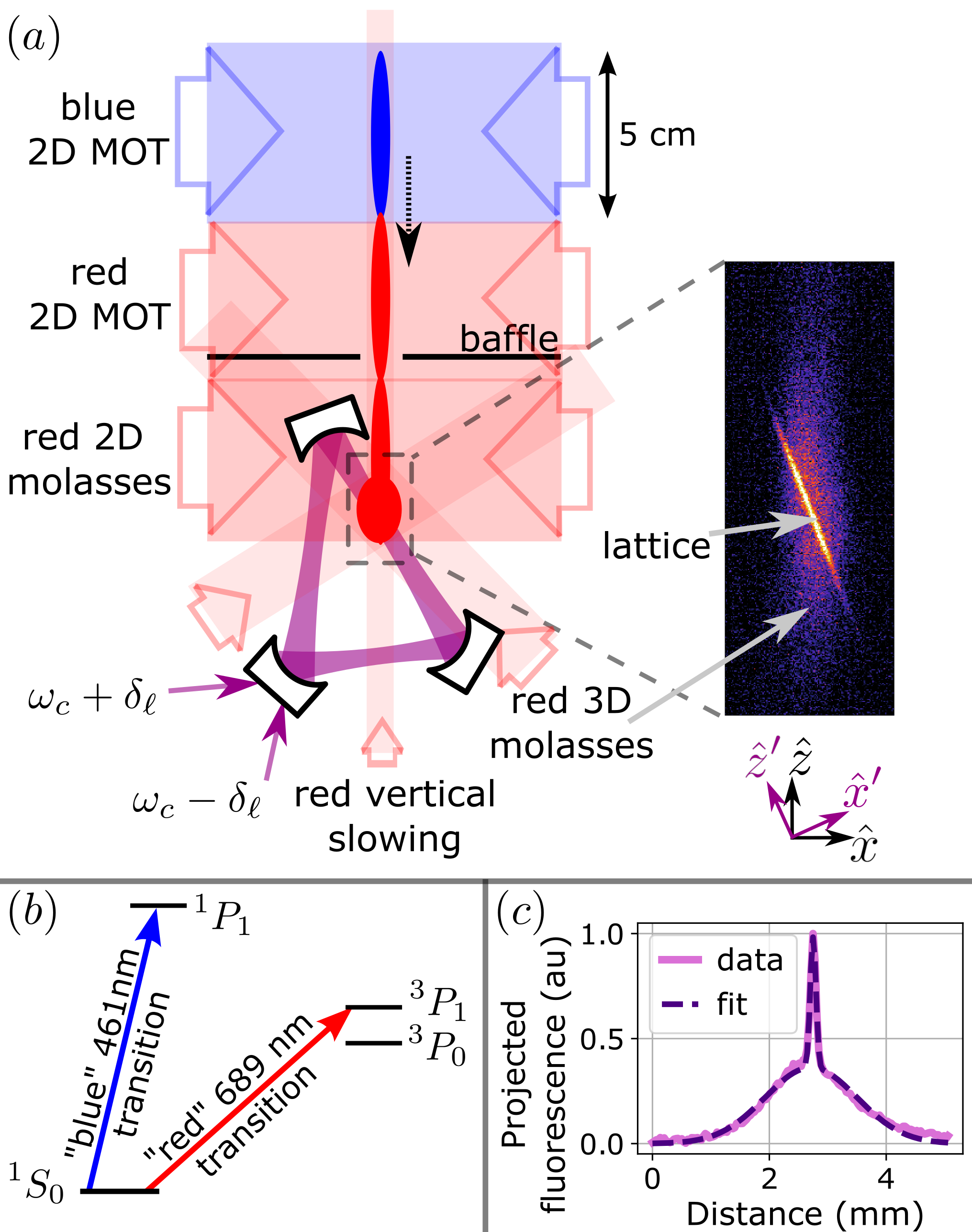}
\caption{Experimental system. (a) Atoms are guided through a series of spatially separated laser cooling and deceleration stages before being captured into a red 3D molasses and loaded into a moving optical lattice supported by the ring cavity.
To form the optical lattice, light is injected into the cavity in the counterclockwise (clockwise) direction at frequency $\wcav \pm \deltal$ to form a lattice travelling at $\deltal$, where $\wcav$ is the cavity resonance frequency.
The inset depicts a single shot image of red florescence from atoms in the red 3D molasses and lattice.
(b) Electronic level scheme of strontium, with the blue and red transitions used for laser cooling. 
(c) Fluorescence of atoms projected onto $\hat{x}'$, the direction perpendicular to the lattice (in pink) and a fit to the sum of two Gaussians (in dashed purple).}
\label{fig:exp}
\end{figure}


To prepare the \sre atoms for being loading into a moving optical lattice inside a ring cavity, they are guided through a series of spatially separated laser cooling and deceleration stages similar in its early steps to \cite{Bennetts2017}. 
We use two optical cooling transitions: the ``blue" dipole allowed 32~MHz linewidth transition with wavelength 461~nm between the ground  $^1$S$_0$ and excited $^1$P$_1$ states, and the ``red" intercombination 7.5~kHz linewidth transition with wavelength 689~nm between the ground  and excited $^3$P$_1$ states, as shown in Fig.~\ref{fig:exp}(b). 
The cooling chain consisting of a Zeeman slower, followed by a blue 2D MOT, a red 2D MOT, a red 2D molasses and a red 3D molasses is shown in Fig.~\ref{fig:exp}(a).
The atoms are transported through the sequential MOT and molasses stages by gravity and radiation pressure forces.

The atomic beam source consists of a high-flux oven and a short permanent magnet-based Zeeman slower operating on the blue transition. 
A blue 2D MOT whose non-confining axis is oriented along gravity, is located 10 cm after the exit of the Zeeman slower (see Fig.~\ref{fig:exp}(a)). 
The quadrupole magnetic field for both the blue and red 2D MOTs is generated by two 3.8~cm permanent magnets centered on the blue 2D MOT, producing a gradient of ~80 G/cm in the center of the blue 2D MOT.
This MOT has a loading rate of $1.1 \times 10^{11}$ $^{88}$Sr atoms/s and a characteristic loading time of 7.0~ms, with a steady-state population of $1.2 \times 10^9$ atoms as measured with fluorescence imaging.

Upon exiting the bottom of the blue 2D MOT, atoms have a radial velocity of 0.5(1) m/s and a downward velocity centered at 6.8(1) m/s, with a root-mean-square (rms) velocity spread of 1.7(2) m/s.
The downward velocity is consistent with radiation pressure forces generated by scattering from atoms in the blue 2D MOT. 
To facilitate capturing the atoms in a red 3D molasses further along the cooling chain we use a vertical red slowing beam which has a $\delta f=-1.6\MHz$ detuning from the $^1\mathrm{S}_0-^3\mathrm{P}_{1,\mathrm{m_J}=-1}$ transition. 
This reduces the downwards velocity of the atoms exiting the blue 2D MOT to 2.6(1)~m/s ($v_{\rm rms}=1.4(1)\mathrm{m/s}$). 

The next cooling stage is a red 2D MOT that operates in a sawtooth wave adiabatic passage (SWAP) configuration \cite{Norcia2018b}, where the frequency is swept over 5.2\MHz\ in a sawtooth ramp across the atomic resonance \cite{Norcia2018b}, with a detuning of the center frequency of $\delta f=-1.4\MHz$ from $^1\mathrm{S}_0-^3\mathrm{P}_{1,-1}$. 
The magnetic field gradient at the center of the red 2D MOT is ~3~G/cm. 

The red 2D MOT is followed by a red 2D molasses ($\delta f=-70\kHz$) and a red 3D molasses ($\delta f=-900\kHz$), which have static frequencies, perform Doppler cooling and form a diffusive trap to enhance the probability of atoms finding the cavity mode.
By the time the atoms reach the cavity mode, with the 3D molasses beams off, the atoms are moving downwards at 1.44(1)~m/s with a root-mean-squared velocity spread of 0.40(2)~m/s, and an atom flux of $5.1 \times 10^8$ atoms/s. 
Overlapped with the cavity mode, the 3D molasses provides cooling in all 3 dimensions and has a lifetime of 260(20)~ms.

Finally the atoms are trapped in an intracavity 1D 813~nm optical lattice formed inside a high finesse cavity. 
The cavity linewidth at the ultranarrow 1.35(3)~mHz 698\nm\ clock transition in $^{87}$Sr \cite{Muniz_mHz} is $\kappa_{\mathrm{clock}} = 2\pi\times 42$~kHz. 
This greatly exceeds the atomic linewidth and places the system in a highly damped, ``superradiant" regime. 
The cavity is 18~cm long, with corresponding free spectral range 1.65~GHz. 
It has a finesse of $4\times 10^4$, and a linewidth of $\kappa = 2\pi\times 50$~kHz at 689\nm\ ($\kappa = 2\pi\times 650$~kHz at 813\nm).

Several design decisions are governed by the requirement that in order to achieve lasing, a cw superradiant laser must be operated in a regime where the collectively enhanced emission rate from the atoms is larger than the atomic decoherence rates, a stringent requirement for ultra-weak transitions.
To increase the collectively enhanced decay rate, we trap the atoms within a high finesse optical cavity, effectively increasing the optical depth of the atomic ensemble. 
To suppress atomic decoherence, we rely on techniques used to provide long coherence times in optical lattice clocks \cite{PhysRevLett.91.173005, Ye1734}, as well as new techniques to provide a compact source with minimal dephasing near the cavity mode. 
Typically, atoms are loaded into a lattice directly from a MOT, guaranteeing low atom temperatures and therefore a high loading efficiency.
However the strong magnetic field gradient inside a MOT would cause inhomogeneous broadening of the atoms due to Zeeman shifts, and the field cannot be switched off temporarily due to the continuous nature of the experiment.
Therefore the two molasses stages are added to allow physical separation between the 2D MOTs and the lattice.
Additionally, a baffle with a 5~mm aperture as a pinhole for the atoms is placed between the red 2D MOT and 2D molasses (shown in Fig.~\ref{fig:exp}(a)). 
The baffle and the interior of the vacuum chamber are painted with low out-gassing black paint to limit the number of stray blue photons in the cavity region.
By laser cooling and confining the atoms along the cavity axis with a magic-wavelength optical lattice, first-order Doppler shifts in the direction of emission are eliminated without imposing large shifts to the lasing transition frequency.

Unlike a two-mirror cavity, a ring cavity can support a moving intracavity optical lattice.
We use this to transport the atoms from the loading region, which is dominated by broadening caused by the cooling and slowing beams, further along the cavity axis to a quieter region with less decoherence.
We chose a three-mirror cavity, rather than a cavity with more mirrors, in order to prevent rotation of the polarization mode in the cavity that is generated by a non-planar configuration \cite{Jaffe2021} and to reduce the mode volume. 
We generate a moving lattice with wavelength $\lambdal=813$~nm moving at $\deltal= 2 \pi v/\lambdal$, where $v$ is the velocity of the lattice with respect to the lab frame, by injecting light into the counterclockwise/clockwise direction at frequency $\wcav \pm \deltal$. 
By choosing the lengths of the cavity arms and radii of curvature of the cavity mirrors to offset the sagittal and tangential waists of the cavity mode, we created a 5~cm region of the main cavity arm where the lattice intensity changes by less than a factor of 2.

Typical lattice trap depths are around $150~\mu$K. 
In steady-state the ratio of number of atoms in the lattice relative to all atoms in the 3D molasses is 0.10(1), as shown in (Fig.~\ref{fig:exp}(c)). 
The lattice loading time for $\deltal=0$ from a fully loaded 3D molasses into an empty lattice is $10(2)$~ms with a maximum atom number of 300,000 atoms. 
This corresponds to a flux of $2.1(3)\cross 10^7$ of atoms/s loaded into the lattice, far exceeding the threshold for superradiance. 


\begin{figure}[!htb]
\includegraphics[width=3.375in]{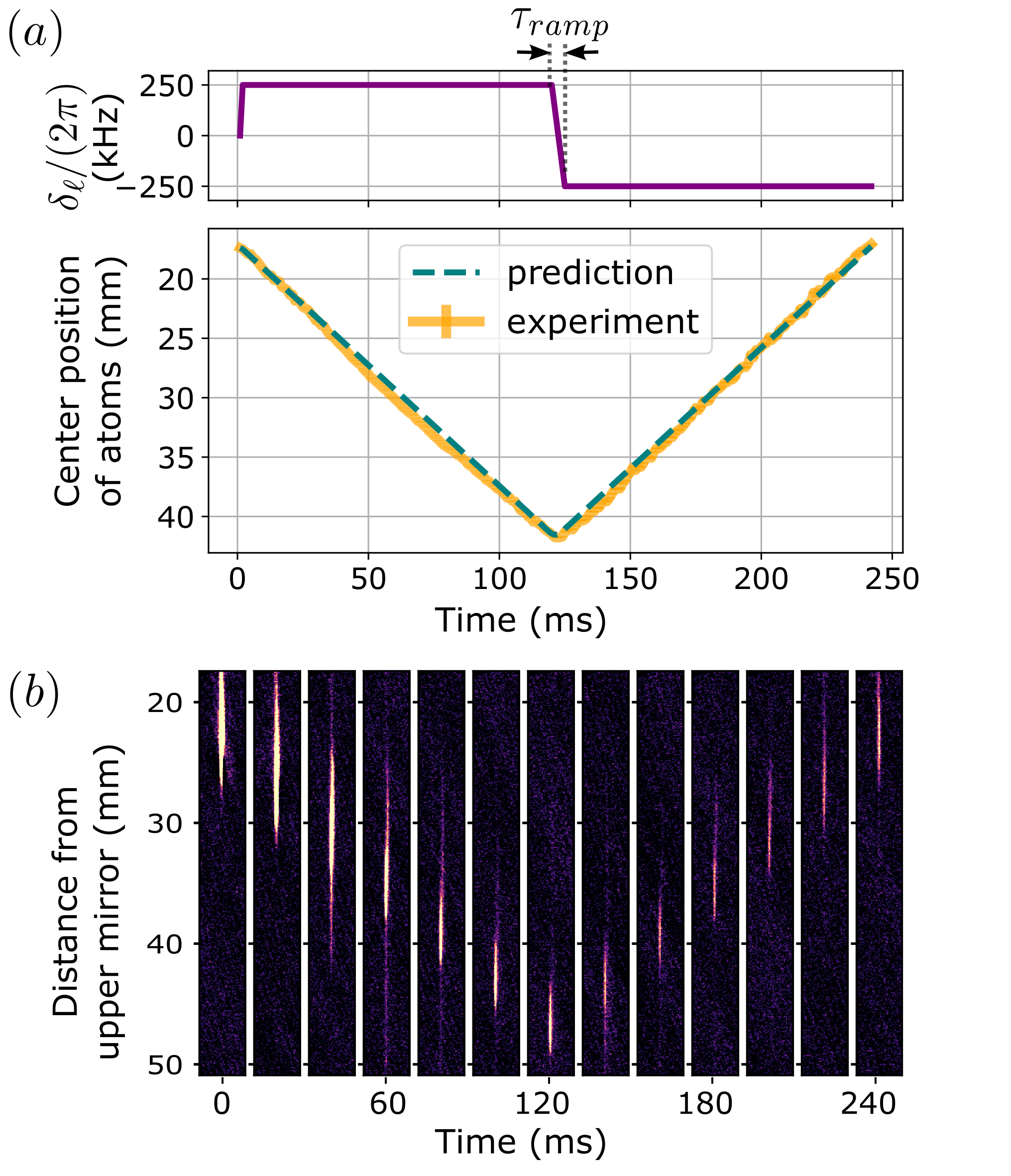}
\caption{Deterministic transport of atoms around the cavity. Atoms are loaded into a stationary lattice ($\deltal=0$), the lattice detuning is linearly ramped to 250 kHz to transport the atoms down the cavity, and then reversed in time $\tramp$ to transport the atoms back up the cavity. For this sequence, (a) shows the lattice detuning and fitted center position of the atomic cloud along the cavity (in yellow) and (b) shows the single shot fluorescence images at selected times. In (a), the  prediction of the center position is the integral of the lattice detuning (shown in dashed teal).}
\label{fig:transport}
\end{figure}

Figure~\ref{fig:transport} demonstrates deterministic transport of the atoms along the cavity axis. 
First, the atoms are loaded into a stationary lattice ($\deltal=0$) and then the lattice detuning is ramped linearly up in time over $\tramp=2.5$~ms to $\deltal=250$~kHz.  We observe that the atoms are transported along the cavity.  After approximately 125~ms of transport, the detuning is then ramped over $\tramp=5$~ms to $\deltal=-250$~kHz, and we see that the atoms are then transported back to the original location.
For this sequence, Fig.~\ref{fig:transport}(a) shows the lattice detuning and fitted center position of the atoms along the cavity and Fig.~\ref{fig:transport}(b) shows the single shot fluorescence images at selected times. 
The prediction of the center position is the integral of the lattice detuning, expressed as a distance.
The ramp time $\tramp$ is chosen to minimize atom loss due to ramping of the lattice detuning $\deltal$.

\begin{figure}[!htb]
\includegraphics[width=3.375in]{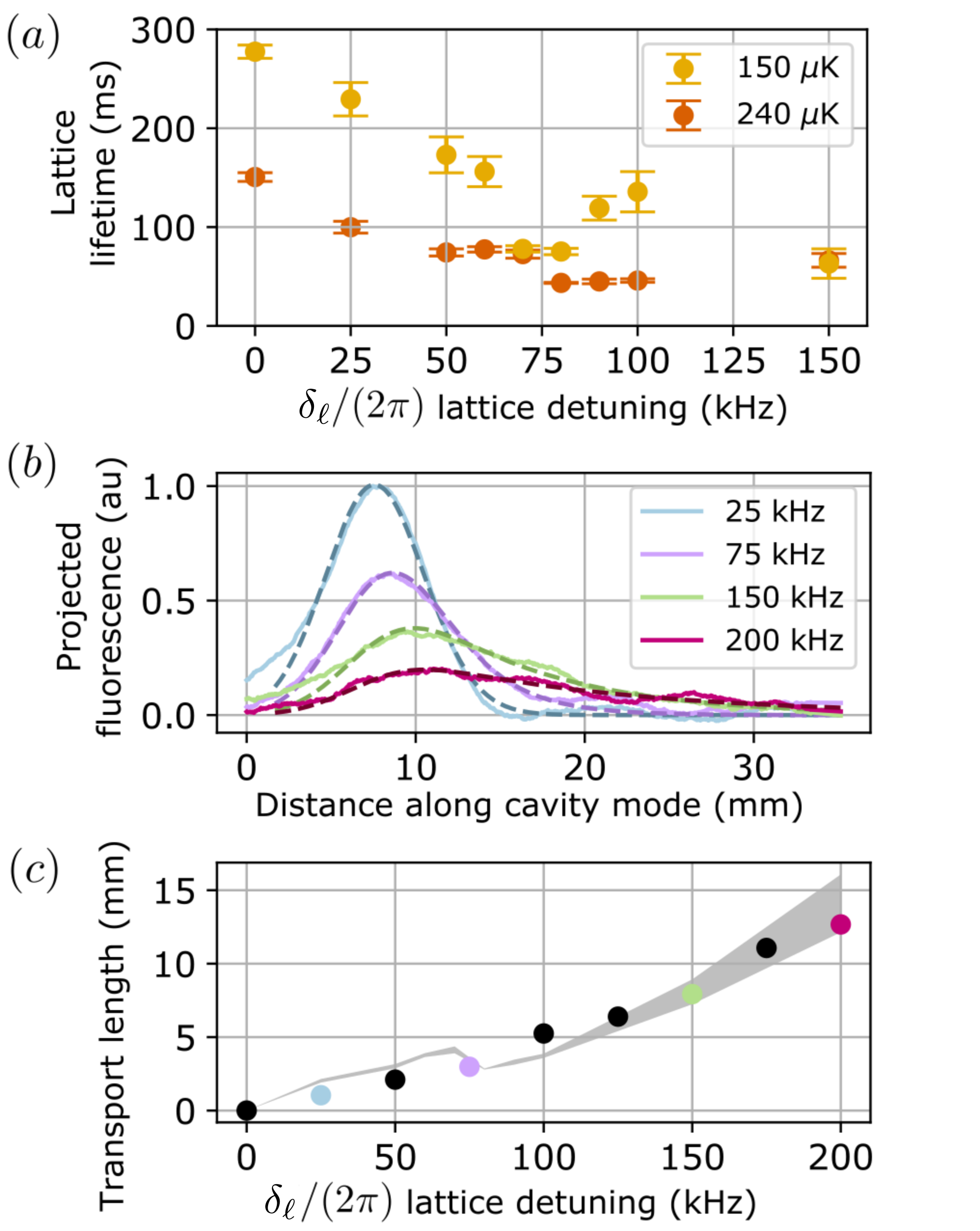}
\caption{(a) Measured lattice lifetime versus lattice detuning \deltal\ for lattice depths $150~\mu$K and 240~$\mu$K are shown in yellow and orange, respectively. (b) Atomic distribution projected onto the cavity mode, after loading atoms into the lattice for 350~ms (solid lines) and fits to Eq.\ref{eq:lt} (dashed), for various cavity detunings. (c) Transport length extracted from the fit in (b) versus lattice detuning. Colored points match the projections plotted in (b) and black points represent additional detunings. The grey shaded region indicates the measured lattice lifetime times the lattice velocity.}
\label{fig:latt}
\end{figure}


\begin{figure}[!htb]
\includegraphics[width=3.375in]{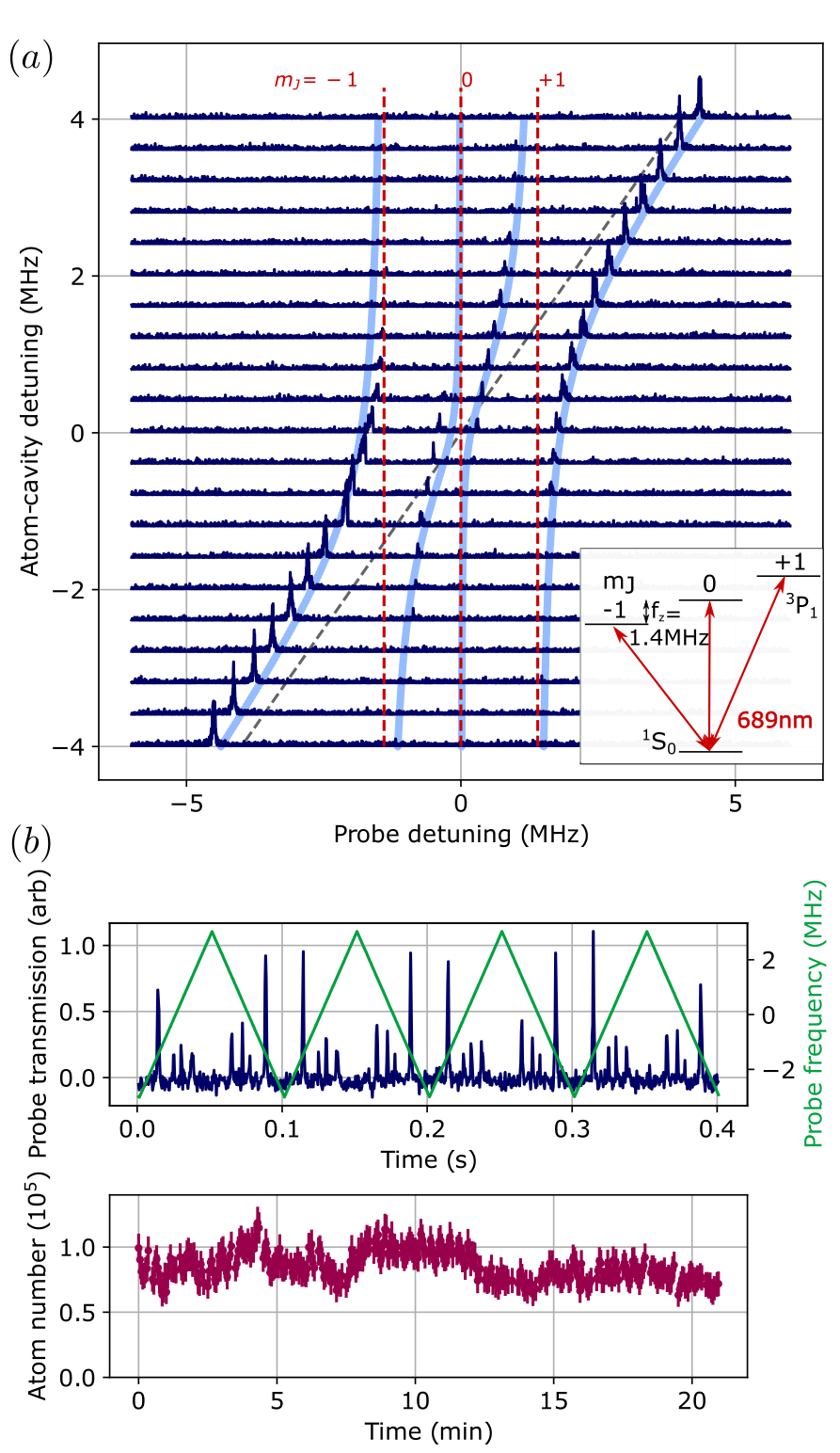}
\caption{Vacuum Rabi splitting: (a) Measured VRS with the cooling lasers switched off 70\us\ before sampling to reduce broadening effects. The probe beam couples to all three $^3P_1$ excited states (red dotted lines), leading to three avoided crossings with the unperturbed cavity mode (grey dotted line). The inset shows the levels involved. The light blue lines show a theory fit to the inferred peak positions of the VRS data. The floated parameters are the magnetic field angle fitted to $30(10)^\circ$ and the atom number $N=6(1)\times 10^{4}$. (b) Continuously sampled VRS over the course of several hours. All cooling beams were on continuously, and the probe frequency was swept in a triangular shape, with the cavity close to resonance with the atomic $^1S_0-^3P_{1,0}$ transition. The bottom plot shows a 20 minute excerpt of continuously measured VRS where each atom number is inferred from the peak splitting across one sweep.}
\label{fig:vrs}
\end{figure}

Figure~\ref{fig:latt} shows atoms loaded directly and continuously into a moving optical lattice. 
The lattice lifetime as a function of the lattice detuning for two different lattice depths is shown in Fig.~\ref{fig:latt}(a). 
The stationary lattice lifetime is limited by scattering of photons of the blue MOT beams. 
The lattice lifetime decreases with increasing trap depth and faster transport velocity.
In a standing wave cavity, the lattice is typically locked on resonance with the cavity, so that there is no first order conversion from laser frequency noise to amplitude noise inside the cavity. 
In our case, we are injecting tones detuned from the cavity resonance so there is a linear conversion from frequency noise to amplitude noise (mitigated by sending in nominally equal power, symmetrically detuned tones about the cavity resonance). 
This higher sensitivity to frequency noise contributes to the lower lattice lifetime at higher lattice detunings.

When atoms are loaded directly into a moving lattice, we observe the formation of a tail of the atom cloud along the cavity axis in the direction of transport that grows with faster transport velocity, see Fig.~\ref{fig:latt}(b).
We can model this atomic density profile $\ncav$ using the atom density profile for the stationary lattice that depends on the Gaussian profile of the red 3D molasses (with standard deviation $\sigma$) as well as the lattice lifetime $\taul$, which maps onto the atoms' extent along the cavity, or transport length, as $\lt = v \cdot \taul$. 
For a moving lattice, the resulting density profile is a convolution of the history of where atoms were loaded in the past and where they have been transported to in the present :
\begin{equation}
\ncav = \int_{0}^{t} R \cdot e^{{\frac{-v(t-t')}{\lt}}} e^{- \frac{(z'-v(t-t'))^2}{2 \sigma^2}} \,dt' 
\label{eq:lt}
\end{equation} 
where $R$ is the loading rate of atoms into the lattice and we started loading atoms into the lattice at time 0 and stopped at time $t$ (in steady-state $t=\infty$). 
Note that this equation can be expressed as error functions.
Figure~\ref{fig:latt}(b) shows the experimentally measured $\ncav$ after loading atoms into the lattice for 350~ms with various lattice detunings, as well as the fit to Eq.(\ref{eq:lt}). 
The fitted $\lt$ is plotted versus lattice detuning in Fig.~\ref{fig:latt}(c). 
The grey shaded region indicates the predicted transport length  (plus or minus  one sigma) using the independently measured lattice lifetime $\taul$ and known transport velocity $v$ as $\lt= \taul \cdot v = \taul \cdot \deltal/(2 \pi) \cdot\lambdal$.  
The model supports that we are continuously loading atoms into a moving intracavity lattice, and indicates that an improved lattice lifetime will substantially increase the rate of atoms delivered into the lower decoherence region of the cavity mode.


To obtain a precise measure of the atom number and demonstrate that we are in the collective strong coupling regime, we have measured the atom-cavity vacuum Rabi splitting (VRS) \cite{Winchester2017}, see Fig.~\ref{fig:vrs}. The cavity frequency is locked at frequency $\wcav$, close to the frequency of the red transition $\watom$ with detuning $\deltac = \wcav-\watom$. 
In typical operation, $\deltac$ is nominally zero ($\left|\deltac\right| \ll \kappa$), but can be varied for characterization purposes. 
The magnetic field at the cavity defines a quantization axis that is tilted $30^\circ$ out of the the plane of the cavity mode and induces a Zeeman splitting of $\omega_z=2 \pi \times 1.4~\mathrm{MHz}$.  
As a result, the probe light with linear polarization perpendicular to the plane of the cavity is a superposition of $\pi$, $\sigma^+$, and $\sigma^-$ polarized light and thus couples to all three Zeeman states of the $^3\mathrm{P}_1$ level.
This leads to the more complex structure of the VRS shown in Fig.~\ref{fig:vrs}(a)~\&~(b).
Several cooling lasers (the 2D molasses, 3D molasses and vertical slowing beams) spatially overlap with the atoms in the lattice region, leading to a slight broadening of the VRS peaks.
To observe the un-broadened VRS, we switch off all cooling lasers $70\us$ before sampling the VRS, corresponding to several lifetimes $\tau=21~\mu\mathrm{s}$ of the excited $^3\mathrm{P}_{1}$ state.
For each probe frequency, we sample the probe light transmitted through the cavity and detected by an single photon counting module (SPCM).
This technique allows us to observe an un-broadened linewidth of the VRS around atom-cavity resonance.
A truly continuous vacuum Rabi splitting sampled over several hours is shown in Fig.~\ref{fig:vrs}(b).
Here the frequency of the probe laser is swept continuously, with all cooling and loading lasers permanently on.

When optimising for atom number up to $1\times10^{6}$ atoms are loaded into the lattice.
This corresponds to a phase space density of $\rho=\frac{N}{V}\lambda_\mathrm{r}^2\lambda_\mathrm{z}=7(2)\cross 10^{-5}$, where $N$ is the total atom number, $V$ is the volume explored by the atoms inside the lattice potential, and the deBroglie wavelength is $\lambda_\mathrm{r/z}=\frac{\hbar\sqrt{2\pi}}{\sqrt{m k_B T_\mathrm{r/z}}}$, with radial/axial temperatures $T_\mathrm{r/z}=10(2)\uK/9(1)\uK$.


In summary, we have demonstrated continuous loading of ultra-cold atoms into a high finesse ring cavity in the strong coupling regime, combined with continuous transport of the atoms along the lattice axis, necessary for transporting the atoms away from cooling lasers to a zone of lower decoherence.
This is an important step towards realising a continuous-wave superradiant laser, an ultra-narrow linewidth active frequency reference with many possible applications for precision measurements.
To achieve cw superradiant lasing the atom flux has to overcome dephasing on the $^3$P$_0$ transition, i.e. the influx of atoms needs to be greater than the number of inverted atoms lost through the lasing process.
Assuming similar excited state dephasing to our previous pulsed experiment \cite{Norcia2016b}, this requires a minimum flux of $N_t/t_w=N_t^2C\gamma/3.5=2.7\times 10^6 \mathrm{atoms/s}$, with threshold atom number $N_t$ \cite{Norcia2016b}, cavity cooperativity $C=0.16$, excited state radiative linewidth $\gamma$ and characteristic emission duration $t_w$ \cite{Gross1982}.
Taking into account atom loss from switching from \sre\!, which has a 10 times higher isotopic abundancy in our source material and can be cooled more efficiently due to its lack of hyperfine structure, to \srs\!, our target species for superradiant lasing due to its narrow linedwith $1.3\mHz$ transition, our achieved atom flux loaded into the lattice reaches this threshold.  In addition, the continuously loaded atoms, homogeneous cavity coupling and high achievable atom numbers demonstrated here give access to new areas of quantum simulation and strongly correlated and non-linear physics regimes.

  
We acknowledge useful input and early discussions with Matthew A. Norcia. We thank Juan Muniz for helpful discussions and construction of a first version of the experiment, and we thank William McGrew and Murray Holland for helpful comments on the manuscript. We acknowledge funding support from DARPA ATN, National Science Foundation under Grant Numbers 1734006 (Physics Frontier Center) and  OMA-2016244 (QLCI), DOE Quantum Systems Accelerator, and NIST. J.R.K.C. acknowledges financial support from NSF GRFP. T.H.Y. acknowledges support by the JILA Visiting Fellows Program.

\bibliography{ThompsonLab.bib}

\end{document}